\documentclass[%
 reprint,
 onecolumn,
 superscriptaddress,
 showpacs,
 aps,
 prstab,
floatfix,
]{revtex4-1}

\usepackage[utf8]{inputenc} 
\usepackage[T1]{fontenc} 
\usepackage{lmodern} 
\usepackage{amsmath}
\usepackage{graphicx}
\graphicspath{{./figures/}}
\DeclareGraphicsExtensions{.pdf}
\usepackage[separate-uncertainty=true, multi-part-units=single]{siunitx}
\usepackage{xspace}
\usepackage{color,soul}

\usepackage[dvipsnames]{xcolor} 

\usepackage[colorinlistoftodos]{todonotes}

\usepackage{enumitem}
\setlist[description]{topsep=0pt,parsep=\parskip,partopsep=0pt,leftmargin=0pt,labelindent=0pt,font=\bfseries\small}

\usepackage{hyperref}
\hypersetup{
    colorlinks=true,
    linkcolor=blue,
    filecolor=blue,      
    urlcolor=blue,
    citecolor=blue,
}
\begin{document}

\title{Demonstration of a compact plasma accelerator powered by laser-accelerated electron beams}

\author{T. Kurz }\thanks{These authors contributed equally to this work}
\affiliation{Helmholtz-Zentrum Dresden--Rossendorf, Bautzner Landstrasse 400, 01328 Dresden, Germany}%
\affiliation{Technische Universität Dresden, 01062 Dresden, Germany}
\author{T. Heinemann}\thanks{These authors contributed equally to this work}
\affiliation{Deutsches Elektronen-Synchrotron DESY, Notkestraße 85, 22607 Hamburg, Germany}%
\affiliation{The Cockcroft Institute, Keckwick Lane, Warrington WA4 4AD, United Kingdom}%
\affiliation{University of Strathclyde, 107 Rottenrow, Glasgow G4 0NG, United Kingdom}
\author{M. F. Gilljohann}
\affiliation{Ludwig--Maximilians--Universit{\"a}t M{\"u}nchen, Am Coulombwall 1, 85748 Garching, Germany}%
\affiliation{Max Planck Institut für Quantenoptik, Hans-Kopfermann-Strasse 1, 85748 Garching , Germany}%
\author{Y. Y. Chang}
\affiliation{Helmholtz-Zentrum Dresden--Rossendorf, Bautzner Landstrasse 400, 01328 Dresden, Germany}%
\author{J. P. Couperus Cabada{\u{g}}}
\affiliation{Helmholtz-Zentrum Dresden--Rossendorf, Bautzner Landstrasse 400, 01328 Dresden, Germany}%
\author{A. Debus}
\affiliation{Helmholtz-Zentrum Dresden--Rossendorf, Bautzner Landstrasse 400, 01328 Dresden, Germany}%
\author{O. Kononenko}
\affiliation{LOA, ENSTA Paris, CNRS, Ecole Polytechnique, Institut Polytechnique de Paris, 91762 Palaiseau, France}%
\author{R. Pausch}
\affiliation{Helmholtz-Zentrum Dresden--Rossendorf, Bautzner Landstrasse 400, 01328 Dresden, Germany}%
\author{S. Sch{\"o}bel}
\affiliation{Helmholtz-Zentrum Dresden--Rossendorf, Bautzner Landstrasse 400, 01328 Dresden, Germany}%
\affiliation{Technische Universität Dresden, 01062 Dresden, Germany}
\author{R. W. Assmann}
\affiliation{Deutsches Elektronen-Synchrotron DESY, Notkestraße 85, 22607 Hamburg, Germany}%
\author{M. Bussmann}
\affiliation{Helmholtz-Zentrum Dresden--Rossendorf, Bautzner Landstrasse 400, 01328 Dresden, Germany}%
\author{H. Ding}
\affiliation{Ludwig--Maximilians--Universit{\"a}t M{\"u}nchen, Am Coulombwall 1, 85748 Garching, Germany}%
\affiliation{Max Planck Institut für Quantenoptik, Hans-Kopfermann-Strasse 1, 85748 Garching , Germany}%
\author{J. G{\"o}tzfried}
\affiliation{Ludwig--Maximilians--Universit{\"a}t M{\"u}nchen, Am Coulombwall 1, 85748 Garching, Germany}%
\affiliation{Max Planck Institut für Quantenoptik, Hans-Kopfermann-Strasse 1, 85748 Garching , Germany}%
\author{A. K{\"o}hler}
\affiliation{Helmholtz-Zentrum Dresden--Rossendorf, Bautzner Landstrasse 400, 01328 Dresden, Germany}%
\author{G. Raj}
\affiliation{LOA, ENSTA Paris, CNRS, Ecole Polytechnique, Institut Polytechnique de Paris, 91762 Palaiseau, France}%
\author{S. Schindler}
\affiliation{Ludwig--Maximilians--Universit{\"a}t M{\"u}nchen, Am Coulombwall 1, 85748 Garching, Germany}%
\affiliation{Max Planck Institut für Quantenoptik, Hans-Kopfermann-Strasse 1, 85748 Garching , Germany}%
\author{K. Steiniger}
\affiliation{Helmholtz-Zentrum Dresden--Rossendorf, Bautzner Landstrasse 400, 01328 Dresden, Germany}%
\author{O. Zarini}
\affiliation{Helmholtz-Zentrum Dresden--Rossendorf, Bautzner Landstrasse 400, 01328 Dresden, Germany}%
\author{S. Corde}
\affiliation{LOA, ENSTA Paris, CNRS, Ecole Polytechnique, Institut Polytechnique de Paris, 91762 Palaiseau, France}%
\author{A. D{\"o}pp}
\affiliation{Ludwig--Maximilians--Universit{\"a}t M{\"u}nchen, Am Coulombwall 1, 85748 Garching, Germany}%
\affiliation{Max Planck Institut für Quantenoptik, Hans-Kopfermann-Strasse 1, 85748 Garching , Germany}%
\author{B. Hidding}
\affiliation{The Cockcroft Institute, Keckwick Lane, Warrington WA4 4AD, United Kingdom}%
\affiliation{University of Strathclyde, 107 Rottenrow, Glasgow G4 0NG, United Kingdom}
\author{S. Karsch}
\affiliation{Ludwig--Maximilians--Universit{\"a}t M{\"u}nchen, Am Coulombwall 1, 85748 Garching, Germany}%
\affiliation{Max Planck Institut für Quantenoptik, Hans-Kopfermann-Strasse 1, 85748 Garching , Germany}%
\author{U. Schramm}
\affiliation{Helmholtz-Zentrum Dresden--Rossendorf, Bautzner Landstrasse 400, 01328 Dresden, Germany}
\affiliation{Technische Universität Dresden, 01062 Dresden, Germany}%
\author{A. Martinez de la Ossa}
\affiliation{Deutsches Elektronen-Synchrotron DESY, Notkestraße 85, 22607 Hamburg, Germany}%
\author{A. Irman}
\affiliation{Helmholtz-Zentrum Dresden--Rossendorf, Bautzner Landstrasse 400, 01328 Dresden, Germany}%


\maketitle


{\bfseries
Plasma wakefield accelerators are capable of sustaining gigavolt-per-centimeter accelerating fields~\cite{Chen1985}, surpassing the electric breakdown threshold in state-of-the-art accelerator modules by 3-4 orders of magnitude. Beam-driven wakefields offer particularly attractive conditions for the generation and acceleration of high-quality beams~\cite{Hidding2012, MartinezdelaOssa2013, Wittig2015}.
However, this scheme relies on kilometer-scale accelerators~\cite{Blumenfeld2007}. 
Here, we report on the demonstration of a millimeter-scale plasma accelerator powered by laser-accelerated electron beams. We showcase the acceleration of electron beams to 130 MeV, consistent with simulations exhibiting accelerating gradients exceeding 100 GV/m.
This miniaturized accelerator is further explored by employing a controlled pair of drive and witness electron bunches, where a fraction of the driver energy is transferred to the accelerated witness through the plasma. Such a hybrid approach allows fundamental studies of beam-driven plasma accelerator concepts at widely accessible high-power laser facilities. It is anticipated to provide compact sources of energetic high-brightness electron beams for quality-demanding applications such as free-electron lasers.}

\vspace{0.2cm}

In beam-driven plasma wakefield accelerators (PWFAs), the space-charge field of an intense and highly relativistic particle beam propagating through a plasma excites a trailing plasma-density wave. 
Following its driver, the associated wakefield enables the acceleration of a witness electron bunch phase-locked to the accelerating field.
For a sufficiently high peak-current drive beam, these plasma wakefields are generated in the blowout regime~\cite{Rosenzweig1991}. In this regime plasma electrons are completely expelled from the driver vicinity, thereby forming a nearly spherical ion cavity~\cite{Lotov2004}. The fields generated by such plasma cavities provide ideal conditions for an accelerator: extreme accelerating gradients are accompanied by strong linear and longitudinally uniform focusing fields allowing for preservation of emittance, a key quality parameter of electron beams. Additionally, PWFAs operating in this regime exhibit stable wakefield formation, which permits persistent beam-loading conditions to yield a low energy spread of the witness beam~\cite{Tzoufras2008}. 
These features, together with the unique prospects for high-quality witness beam injection~\cite{Hidding2012, MartinezdelaOssa2013, Wittig2015}, open a door to a new generation of monochromatic, ultra-bright beams for applications that demand high beam quality~\cite{Gruener2007}. 

So far, the limited availability of high peak-current drive beams has constrained the development of PWFAs to only a few dedicated facilities. 
Nowadays, compact laser-driven wakefield accelerators (LWFAs)~\cite{Downer2018}, hosted in many high-power laser facilities worldwide, can deliver GeV-class electron beams~\cite{Gonsalves2019} at peak-currents exceeding ten kiloampere~\cite{Li2017,Couperus2017}.
In conjunction with an inherently short duration of a few femtoseconds~\cite{Lundh2011}, such LWFA beams constitute ideal drivers for PWFAs~\cite{Hidding2010,MartinezdelaOssa2015} at plasma densities above $\SI{e18}{\per\cubic\centi\metre}$,
where accelerating gradients of 100 GV/m can be generated~\cite{Corde2016}.
Even with a sizeable energy spread and emittance, LWFA beams provide attractive attributes for improved resilience to driver instabilities~\cite{Mehrling2017, MartinezdelaOssa2018}.
Thus, utilizing LWFA beams as PWFA drivers in a staged LWFA-driven PWFA (LPWFA) harnesses the unique advantages of each plasma acceleration method in a compact geometry~\cite{MartinezdelaOssa2019}. 
Specifically, the capability of LWFAs to deliver high peak-current drive beams can be combined with the potential of PWFAs to ultimately deliver electron beams of superior quality.
Additionally, LPWFAs benefit from inherent laser-to-beam synchronization, a key advantage for advanced injection methods employing assisting laser pulses~\cite{Deng2019}.

\begin{figure}[t!]
  \includegraphics[width=1.0\columnwidth]{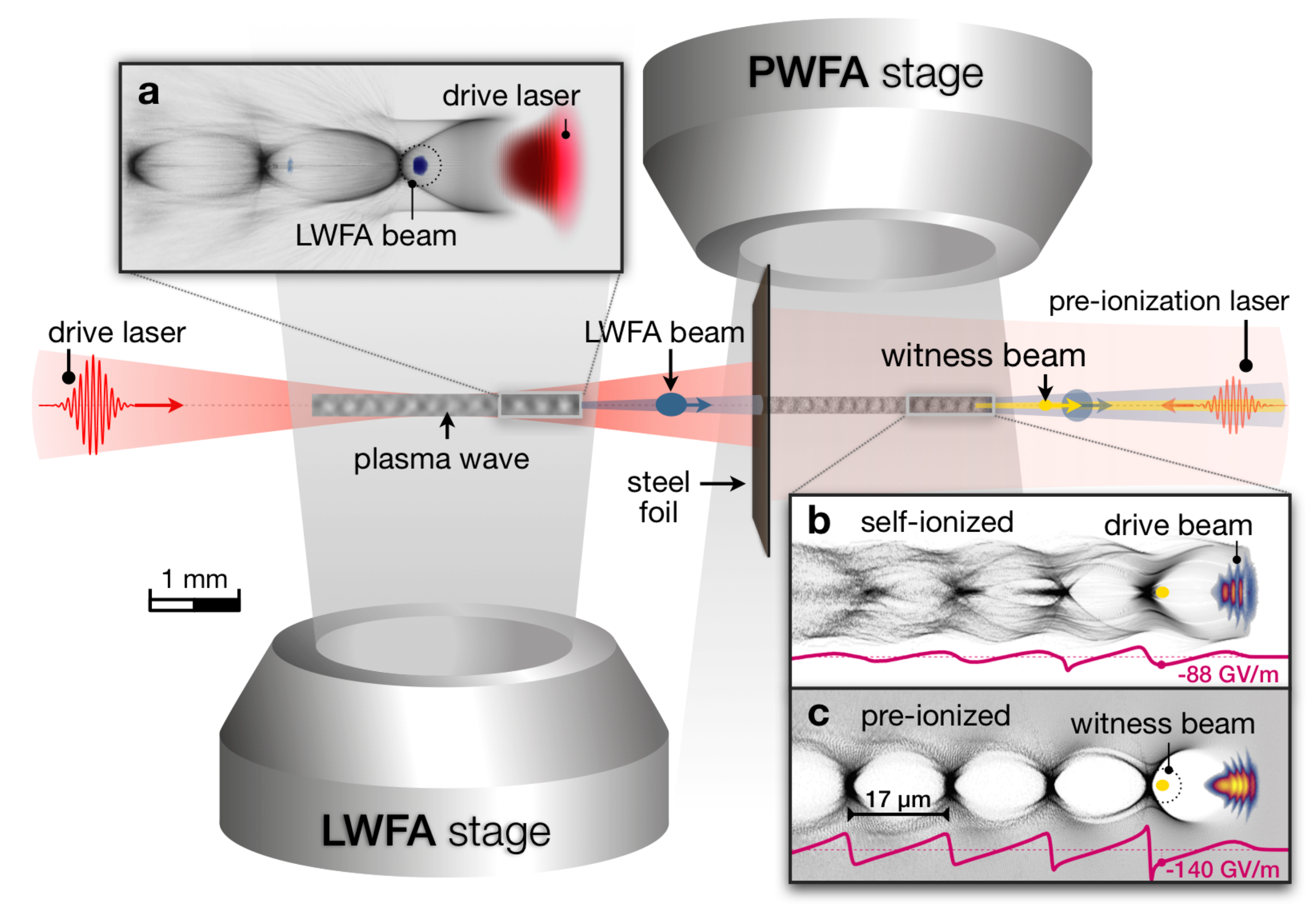}
  \caption{
    \textbf{Schematic overview of the experiment.} Two consecutive gas jets form the basis of an LWFA-driven PWFA. From left to right, a high-intensity laser pulse (\textit{red}) drives an LWFA in the first stage \textbf{(a)}, generating a high peak-current electron beam (\textit{blue}). The spent LWFA laser is reflected by a thin steel foil, whereas the electron beam propagates into the second stage, acting as the PWFA driver. In the PWFA stage, a witness beam (\textit{yellow}) is accelerated. Additionally, a counter-propagating low-power laser pulse can be applied for generating a pre-formed plasma channel in the PWFA stage prior to the drive beam arrival. Subfigure \textbf{(b)} and \textbf{(c)} show plasma wakefields obtained from simulations using the code OSIRIS considering only the interaction of the drive beam with the second gas jet at the self-ionized and pre-ionized regime, respectively, with the purple line representing the longitudinal electric field on axis.}
  \label{fig:setup}
\end{figure}

Here, we report on two complementary experimental implementations of an LPWFA, performed at $100~\mathrm{TW}$-class, short-pulse laser facilities: the DRACO laser at the Helmholtz-Zentrum Dresden--Rossendorf and the ATLAS laser at the Ludwig-Maximilians-Universit\"at M\"unchen. 
In contrast to initial experimental works exploring a transition from laser- to beam-driven modes in single or coupled stages~\cite{Corde2011, Masson-Laborde2014}, we employ two individual gas jets operating as the LWFA and PWFA stages, respectively. 
This enables independent control and optimization of each stage and unambiguous distinction of beam-driven from laser-driven acceleration. 

In order to demonstrate witness beam acceleration in a high-gradient PWFA, the LWFA stage is optimized to generate high peak-current drive electron beams~\cite{Couperus2017}.  For this purpose, the self-truncated ionization-induced injection scheme~\cite{Zeng2014} is deployed. As sketched in Fig.~\ref{fig:setup}, the LWFA stage consists of a 3 mm-long helium gas jet doped with 3$\%$ nitrogen (see Methods). 
The PWFA stage is formed by a 3 mm-long hydrogen gas jet doped with 10$\%$ helium, which is located directly behind the first stage avoiding any vacuum gap in between (see Methods).
A \SI{12.5}{\micro \meter} thick steel foil is positioned at the entrance of the PWFA section to reflect the spent laser pulse, while the electron beam passes through the foil and drives a purely beam-driven wakefield. 
In this setup, the PWFA stage can be either self-ionized by the space-charge field of the electron drive beam or pre-ionized by a dedicated counter-propagating laser pulse (see Methods). 
A synchronized few-cycle laser pulse provides a side view of the corresponding plasma waves in the PWFA stage via shadowgraphy (see Methods).

\begin{figure*}
  \centering  \includegraphics[width=1.0\columnwidth]{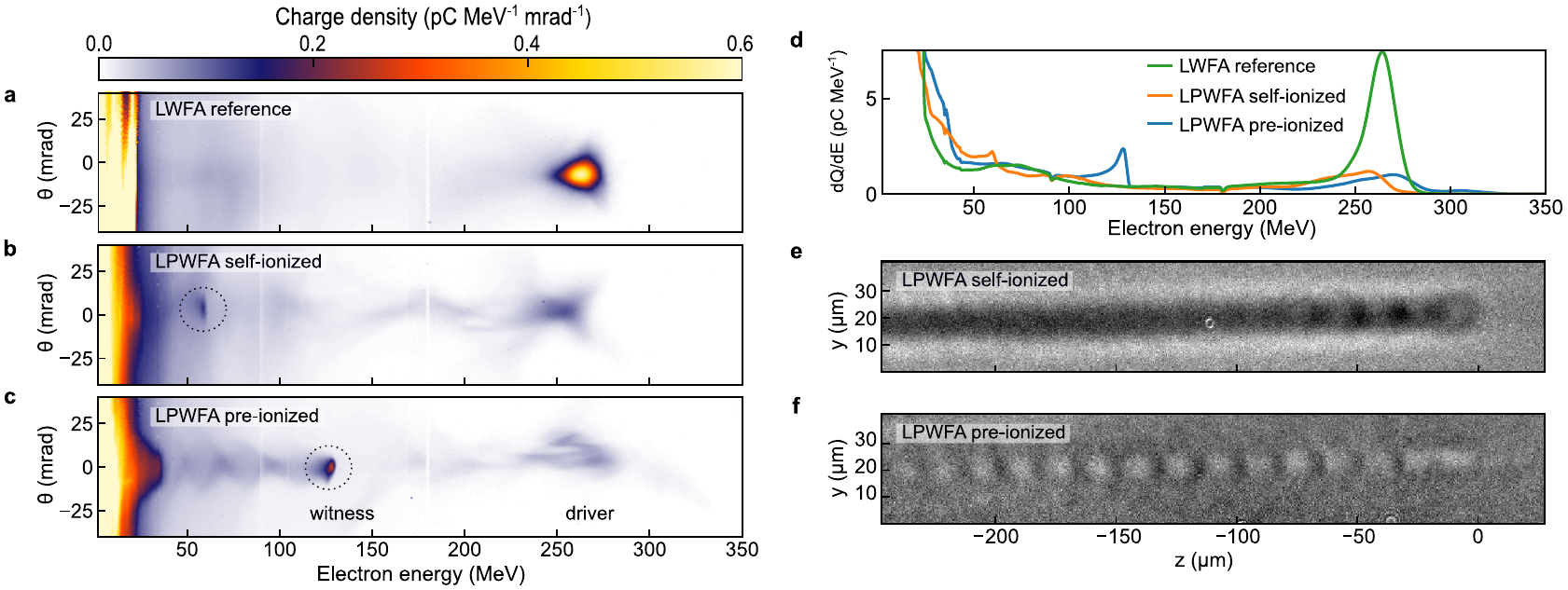}
  \caption{\textbf{Representative electron spectra and plasma wave shadowgrams.} \textbf{(a)} Energy spectrum of LWFA electrons transmitted through the steel foil without operating the PWFA stage and $\theta$ representing the divergence. \textbf{(b)} LPWFA spectrum without pre-ionizing laser. \textbf{(c)} LPWFA spectrum with the PWFA stage ionized prior to the drive beam arrival. \textbf{(d)} Charge distribution integrated over $\pm15$ mrad divergence: \textit{green line} corresponds to \textbf{a}, \textit{orange} to \textbf{b} and \textit{blue} to \textbf{c}. Additional spectra used for determining the statistical parameters can be found in Supplementary Fig.~1, with the measured beam parameters from all shots summarized in Supplementary Table 1. \textbf{(e)} Plasma wakefield shadowgram at the center of the PWFA stage, with the drive beam propagating to the right and ionizing the gas by means of its electric field (self-ionized case). The ionized channel (dark region) and several plasma wakefield oscillations directly behind the driver are visible. \textbf{(f)} Corresponding shadowgram of a plasma wakefield in the pre-ionized case.}
  \label{fig:waterfall}
\end{figure*}

As a reference set, we first recorded shots with only the LWFA stage in operation with the steel foil on position. 
The resulting electron spectra exhibit typical narrow-band peaks with a $262\pm9~\mathrm{MeV}$ shot-averaged mean energy ($\bar{E}_\Delta$), a full-width at half-maximum (FWHM) bandwidth ($\Delta E$) of $ 24\pm 4~\mathrm{MeV}$ and an FWHM-integrated charge ($Q_\Delta$) of $104 \pm 12~\mathrm{pC}$. 
The corresponding shot-averaged spectral charge density, defined as $\mathcal{S}_\Delta = Q_\Delta/(\Delta E/\bar{E}_\Delta)$, is $11.4 \pm 2.1~\mathrm{pC/\%}$.  
A representative reference spectrum is shown in Fig.~\ref{fig:waterfall}(a).  
A considerable amount of charge is also observed at low energies up to $\sim 30~\mathrm{MeV}$, attributed to electrons originating from the plasma density down-ramp transition at the end of the LWFA stage, as also observed in~\cite{Couperus2017}. 
The interaction between the LWFA electrons and the foil increases the divergence of the LWFA beam by $50\%$ (see Supplementary Fig.~1) but does not significantly compromise its ability to drive plasma waves, due to the close proximity between the foil and the PWFA stage. 

With both jets turned on, the PWFA stage is first operated without pre-ionization. 
A clear signature of the drive beam interaction with the second stage is observed, as exemplified in Fig.~\ref{fig:waterfall}(b). 
The shot-averaged spectral charge density decreases to one third of the value obtained for the LWFA reference shots, due to spectral broadening and charge loss (see Supplementary Table~1), as also seen in~\cite{Chou2016}. 
This implies that the drive beam ionizes the ambient gas and transfers a fraction of its energy into the plasma.
This hypothesis is confirmed by the shadowgraphy images recorded inside the PWFA stage, depicted in Fig.~\ref{fig:waterfall}(e), which show a narrow plasma filament inside the otherwise neutral gas along the drive beam propagation axis. 
A few oscillation periods of a plasma wave are observed, clearly demonstrating that the space charge field of the drive beam is sufficiently high to not only ionize the gas but also to excite wakefields. 
In this self-ionized regime only a fraction of the drive beam participates in plasma wakefield formation, resulting in a comparatively weak accelerating gradient, as also confirmed by the simulation shown in Fig.~\ref{fig:setup}(b). Importantly, Fig.~\ref{fig:waterfall}(b) shows a distinct signature of an accelerated witness beam at an energy of about 60 MeV. 

In contrast to the self-ionized regime, creating a pre-formed plasma environment allows the whole drive beam to contribute to the plasma wakefield formation, thus transferring more energy to the plasma and driving a larger amplitude wakefield, as illustrated in simulation Fig.~\ref{fig:setup}(c). 
This increased interaction with the plasma consequently results in a stronger drive beam degradation, as observed in Fig.~\ref{fig:waterfall}(c) (see Supplementary Fig.~4 for supporting simulations). The spectral charge density with respect to the LWFA reference remains about $17\%$ on average, approximately half of the value measured in the self-ionized case. The associated shadowgram shown in Fig.~\ref{fig:waterfall}(f) reveals a more pronounced plasma wakefield structure extending beyond \num{10} subsequent cavities, supporting observations reported in ~\cite{Gilljohann2019}.
As the primary finding, the witness beam energy is significantly increased up to about 130 MeV, about twice the energy observed when operating without the pre-formed plasma channel (see Supplementary Fig.~2 and Fig.~3).
Because the pre-ionizing laser pulse only influences the second stage behind the foil, where no LWFA laser is present, this increase in witness energy must therefore be attributed to the larger amplitude of the beam-driven plasma wakefield.
Assuming an acceleration distance of $\sim$1.5 mm, from the foil until the end of the plasma density plateau (see Methods), we thus estimate an effective accelerating gradient $\sim$50 GV/m higher than in the self-ionized regime. 
Even without precise knowledge of the witness injection energy, this difference represents a conservative lower limit for the true accelerating gradient. 
Nevertheless, such field strength is already comparable to what has been previously achieved at RF-based PWFA experiments~\cite{Blumenfeld2007, Litos2014, Corde2016}.
In agreement with the observation of beam-driven plasma wakefields and the consistent drive beam degradation, this finding provides a conclusive evidence of witness beam acceleration in an LPWFA.

\begin{figure}[htb]
  \includegraphics[width=1.0\columnwidth]{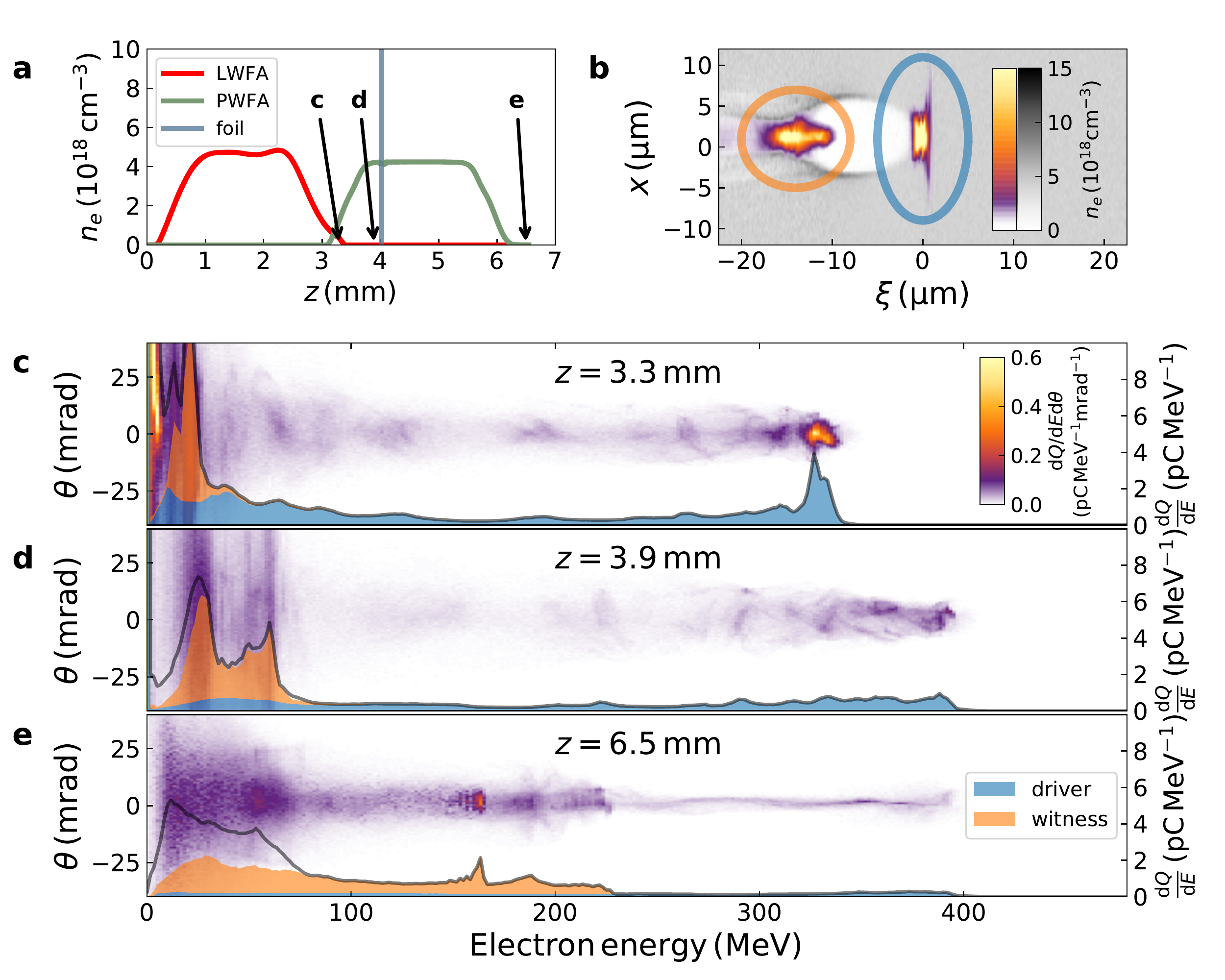}
  \caption{
    \textbf{Results from start-to-end PIC simulations for the pre-ionized case.} \textbf{(a)} Plasma density profile in the simulation, according to the actual experimental geometry and measured density profiles for both gas jets. The drive laser pulse propagates to the right. \textbf{(b)} Electron density distribution in the PWFA stage, 1.5 mm after the foil. The LWFA electron beam (\textit{circled in blue}) excites a plasma wakefield, with the witness beam (\textit{circled in orange}) accelerating at the back of the first cavity. $\xi=z-ct$ represents the co-moving coordinate parallel to the drive beam propagation direction z. The speed of light and time are denoted by c and t, respectively. \textbf{(c)} Simulated electron spectrum after the LWFA stage (z=3.3mm), showing a similar quasi-monoenergetic distribution of the drive beam as measured in experiments. \textbf{(d)} Just before the foil (z=3.9mm), the wakefields driven by the spent laser pulse lead to an energy gain of the LWFA beam. \textbf{(e)} After the PWFA stage (z=6.5mm) a strong degradation of the driver along with energy gain of the witness beam (\textit{orange}) is observed.  }
  \label{fig:PIC-sim}
\end{figure}

The origin of the witness beams can be addressed by quantitative start-to-end simulations. As illustrated in Fig.~\ref{fig:PIC-sim}, they suggest that the observed witness beams may consist of electrons generated at the density down-ramp of the first gas jet, which get trapped and accelerated in the PWFA stage. Three-dimensional particle-in-cell (PIC) simulations using the PIConGPU code were performed for the experimental conditions of the pre-ionized LPWFA (see Methods). 
The simulation setup is shown in Fig.~\ref{fig:PIC-sim}(a). 
At the end of the LWFA stage, the simulated electron spectrum, depicted in Fig.~\ref{fig:PIC-sim}(c), shows parameters consistent with the experiment and reproduces the generation of the low-energy background electrons. 
At the region in front of the foil, the spent laser pulse is still strong enough to drive a plasma wakefield which further accelerates the LWFA electrons. 
The resulting energy gain is visible in the electron spectrum presented in Fig.~\ref{fig:PIC-sim}(d).  
After the laser is reflected by the foil, the transition to purely beam-driven acceleration occurs. 
Fig.~\ref{fig:PIC-sim}(b) shows the nonlinear plasma wakefield driven by the LWFA electron beam behind the foil. 
There, some of the low-energy background electrons are captured by the beam-driven wakefield and are further accelerated to form a witness beam, as shown in Fig.~\ref{fig:PIC-sim}(e). 
The simulations reveal that the witness beam experiences accelerating gradients exceeding $100~\mathrm{GV/m}$ at the electron density plateau in the PWFA stage. 

\begin{figure}[t!]
  \includegraphics[width=1.0\columnwidth]{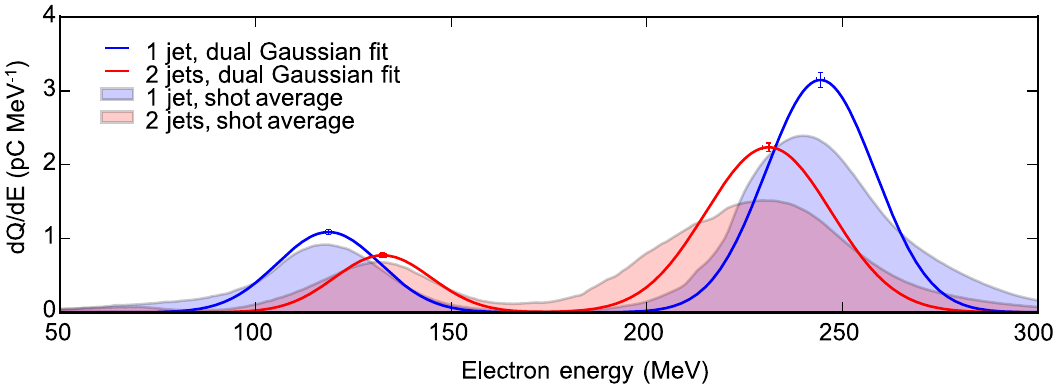}
  \caption{
    \textbf{LPWFA using a drive-witness bunch pair.} Light-blue shaded area: 197-shot-averaged LWFA reference spectra, showing a dual-energy bunch distribution. The drive bunch possesses a higher energy than the witness. To identify the energy peaks of such a bunch pair, a double Gaussian fit was performed on the electron spectrum for each shot. The solid blue line represents the double Gaussian with the shot-averaged fit parameters for the reference set. The error bars indicate the standard error of the average peak position. The Red-light shaded area: 165-shot-averaged electron spectra with both the LWFA and PWFA stages turned on and the solid red line represents the corresponding double Gaussian fit. Note that the deviation of the fits from the measured data is mainly caused by the broadening of the average spectra due to shot-to-shot fluctuations. A systematic deceleration of the drive beam and an acceleration of the witness is clearly observed.}
  \label{fig:LMU-initialresult}
\end{figure}

Demonstrating the capabilities of LPWFAs to accelerate witness beams serves as the basis for various techniques of controlled injection.
Taken from established concepts developed in RF-based PWFA facilities~\cite{Litos2014}, this control includes the preparation of pairs of distinct drive and witness bunches with known temporal separation~\cite{Hidding2010,Wenz2019}. 
Here, a drive-witness bunch pair is produced in the LWFA stage by optimizing the shock-front injection scheme~\cite{Buck2013} such that multiple plasma cavities can be filled. 
A bunch trailing in the second cavity, separated from the driver by approximately one plasma wavelength, represents the witness for the PWFA stage (see Supplementary Fig.~6). 
This method furthermore ensures a fixed energy ratio of the bunch pair, with well-separated energy peaks observed in the averaged LWFA spectrum shown in Fig.~\ref{fig:LMU-initialresult}. 
The average single-shot energy distribution shows a mean energy of $244 \pm 2~\mathrm{MeV}$ for the driver and $119 \pm 1~\mathrm{MeV}$ for the witness. 
In this experiment, no laser blocker foil was used but the distance between the stages was increased to 6 mm, such that the laser intensity is substantially reduced due to its natural diffraction. 
Thus, the laser at the entrance of the second gas jet takes the role of the pre-ionizing laser, but does not drive significant wakefields. 
With a small 1-mrad (FWHM) divergence, the LWFA bunches maintain a high charge density, resulting in the creation of a beam-driven plasma wakefield in the second stage~\cite{Gilljohann2019}. 
In order to position the witness at an accelerating phase of the wakefield and to reduce the spectral overlap between both bunches after the PWFA stage, the second gas jet is operated at a relatively low density of approximately one third of that of the LWFA stage (see Supplementary Fig.~5). 
In concordance with the experiment presented previously, a degradation of the drive bunch in charge and energy spread is observed for shots with both jets active. Consequently, the measured electron spectra show a clear signal of drive bunch deceleration with simultaneous acceleration of the witness bunch to $133 \pm 1~\mathrm{MeV}$. The energy gain of the witness together with the consistent drive bunch deceleration conclusively demonstrate the onset of a dual-beam LPWFA.

In conclusion, acceleration of witness electron beams is demonstrated in a high-gradient PWFA driven by intense laser-accelerated electron beams. Successful operation of this scheme is shown using controlled pairs of drive and witness bunches prepared in the LWFA stage.
Our results substantiate that the LPWFA scenario can be implemented into typical LWFA facilities, which makes PWFA research and applications more accessible. 
This paves the way for a wide range of novel hybrid plasma accelerator systems, such as a PWFA energy booster stage~\cite{Hidding2010} based on controlled and tunable drive-witness pair production in the LWFA stage~\cite{Wenz2019}. 
Furthermore, the high wakefield amplitudes and the inherent laser-to-beam synchronization, unique to the LPWFA scheme, will allow the implementation of advanced internal injection schemes, specifically developed for generating ultra high-brightness beams with unprecedentedly low emittance and energy spread~\cite{Hidding2012, MartinezdelaOssa2013, Wittig2015}. 
Therefore future implementations of LPWFAs can be used as beam brightness and energy transformers, delivering high-quality beams at multi-GeV energies while maintaining a compact setup~\cite{MartinezdelaOssa2019}. 
Such electron beams would be compliant with beam-quality-demanding light sources such as compact free-electron lasers~\cite{Gruener2007} and the long-term goal of plasma-based colliders with high luminosity~\cite{Edda2019}.

\section*{Methods}

\subsection{Laser system}

The high-gradient LPWFA experiment was performed at the DRACO Ti:Sa chirped pulse amplification laser system at the Helmholtz-Zentrum Dresden--Rossendorf (HZDR)~\cite{Schramm2017}.
The system delivers pulses of 30 fs (FWHM) duration at 800 nm central wavelength. 
In this work, a pulse energy of 1.7 J was applied on target, after a small energy extraction of about 21 mJ 
for the counter-propagating pre-ionization laser and the few-cycle probe pulse. 
The remaining part of the pulse was focused by an $F/20$ off-axis parabolic mirror onto the LWFA stage.
The focal spot profile was optimized to a nearly diffraction-limited far-field by performing a wavefront correction on the laser near-field with a wavefront sensor (SID4-Phasics) in closed loop with a deformable mirror, resulting in a FWHM spot size of \SI{19.5}{\micro \meter} as measured at the vacuum target focus position. 
The estimated peak intensity $I_0$ equals \SI{1.0e19}{\watt \per \square \centi \meter}, corresponding to a normalized vector potential
$a_0 \approx 2.1$. 
The spectral shape was measured with spectral-phase interferometry for a direct electric field reconstruction (SPIDER-A.P.E) in conjunction with a self-referenced spectral interferometry (WIZZLER-Fastlite). An acousto-optic programmable dispersive filter (DAZZLER-Fastlite) was used in a closed loop for the correction of any dispersion mismatch between stretcher, compressor, dispersive materials and beamline optics.
During operation, online diagnostics for far-field, near-field and temporal stability situated at the experimental area were used to ensure stable shot-to-shot performance of the laser.    

Similar conditions and techniques were employed at the ATLAS Ti:Sa laser system at Ludwig-Maximilians-Universit\"at (LMU) M\"unchen, where the dual-bunch LPWFA experiment was performed using a pulse energy of 2.5 J on target with an FWHM duration of 28 fs (80 TW) at 800 nm central wavelength. The pulses were focused on the first gas target to a \SI{30}{\micro \meter}-FWHM focal spot and a corresponding peak intensity of \SI{6.9e18}{\watt\per\square\centi\meter} reaching $a_0 \approx 1.8$. 

\subsection{Laser-wakefield acceleration stage}
In the high-gradient LPWFA experiment, the laser-wakefield acceleration stage was operated in a tailored regime of the self-truncated ionization-induced injection scheme~\cite{Zeng2014}, generating high-charge electron beams. This scheme employs a low ionization threshold (LIT) gas as the plasma medium doped with a small fraction of a high ionization threshold (HIT) gas.
The inner electrons of the HIT gas are ionized and subsequently injected only in the vicinity of the intensity peak of the laser pulse located at the front of the plasma bubble. The truncation, which limits the electron injection time, is caused by the nonlinear evolution of the laser pulse in the plasma and the correlated evolution of the plasma cavity. A more detailed description of this regime can be found in \cite{Irman2018}. 

The plasma medium was provided by a 3 mm supersonic de Laval nozzle (Mach 10.4) attached on a fast valve (Parker 9-series) operated using a pre-mixture of helium (97\%) and nitrogen (3\%) acting as the LIT and the HIT gas species, respectively. 
Before the experiment, the gas profile was characterized by a dedicated tomographic interferometry setup~\cite{Couperus2016}, yielding a flat top region of $\sim$1.6 mm with density ramps of $\sim$0.6 mm on both sides along the laser propagation axis. 
The gas pressure was set to 14-16 bar, resulting in a plasma density of \SIrange{4}{4.5e18}{\per\cubic\centi\meter}. 

\subsection{Plasma-wakefield acceleration stage}
For the PWFA stage, a gas nozzle with a geometry identical to the one in the LWFA stage was used, operated by a pre-mixture of hydrogen (90\%) and helium (10\%). 
The stage was operated at a plasma density of \SIrange{3.5}{4.2e18}{\per\cubic\centi\meter} assuming the full ionization of hydrogen and the first level of helium. 
The use of a gas mixture was intended to in principle enable ionization-based injection schemes~\cite{Oz2007,MartinezdelaOssa2013}. The PWFA nozzle was oriented at a ninety degree angle with respect to the LWFA nozzle in order to minimize turbulence of jet flow between both stages. 
The PWFA stage was positioned after the LWFA stage such that the LWFA gas-jet downramp and the PWFA gas-jet upramp were directly adjoined without any vacuum gap in between, see the gas profile in Fig.~\ref{fig:PIC-sim}(a).

\subsection{Laser-blocker foil}
A \SI{12.5}{\micro\meter} thick steel foil was used to reflect the spent LWFA driver-laser entering the PWFA stage. Mounted on a rotational disc, the foil was refreshed for each shot. The foil position with respect to both jets could be adjusted. In the work presented here it was positioned at $\sim$\SI{700}{\micro\meter} upstream from the center of the second gas jet. 

\subsection{Pre-ionizing laser}
Optionally, the PWFA stage gas-medium could be ionized prior to the drive beam arrival. 
This is achieved by a $\sim$\SI{20}{\milli \joule} laser counter-propagating under a shallow angle through the PWFA stage, about 1 ps before the arrival of the LWFA beam. 
A curved mirror with a focal length of $f = \SI{1}{\meter}$ was used to focus the laser to a spot size of $\sim$\SI{120}{\micro \meter} (FWHM) corresponding to a focal peak intensity of \SI[per-mode = symbol]{4e15}{\watt \per \square \centi \meter}, which is well above the ionization threshold for hydrogen and helium. The ionization laser is prevented from entering the LWFA stage by the steel foil. 
However, the intensity of the pre-ionization laser was sufficiently low to not compromise the integrity of the blocker foil.

\subsection{Few-cycle probe laser}
A few-cycle laser pulse was used for ultrafast probing of the PWFA stage by recording shadowgrams of the plasma waves. The probe generation setup consists of a \SI{1.0}{\metre}-long hollow core fiber filled with \SI{2.0}{\bar} of neon, which was seeded using a 1 mJ beam picked up from the main laser pulse, thus inherently synchronized. After this laser pulse was spectrally broadened inside the fiber, it was compressed using chirped mirrors to a pulse length of \SI{9.2}{\femto\second} measured by a spectral-phase interferometry for a direct electric-field reconstruction (SPIDER-A.P.E). The output energy was measured to be \SI{0.4}{\milli\joule} at a beam diameter of \SI{7}{\milli\metre}.
This probe beam was directed to the center of the PWFA stage, transversely illuminating the plasma wakefield which was imaged by a long working distance objective onto a 14 bit CCD camera with a spatial resolution of \SI{0.46}{\micro\metre} per pixel. A similar setup was used in~\cite{Gilljohann2019}.

\subsection{Electron beam characterization}
The electron beam spectral distributions were determined using a \SI{0.4}{\meter} long permanent-magnet dispersive dipole spectrometer with a magnetic field strength of \SI{0.9}{\tesla}. 
Phosphor-based scintillating screens (Konica Minolta OG 400), imaged to 12 bit CCD cameras, were positioned such that the energy resolution is optimized with point-to-point imaging up to 200 MeV. At higher energies the readout error is dominated by the beam pointing error, with a readout uncertainty of \SI{-1.2/+1.6}{\percent} at \SI{300}{\mega \electronvolt} and  \SI{-2.5/+3.1}{\percent} at \SI{400}{\mega \electronvolt} for a 6 mrad pointing error~\cite{Schramm2017}. The overall detection range is \SIrange[range-phrase=--, range-units = single]{2}{550}{\mega \electronvolt}.

In order to deduce the beam charge-energy distribution, the absolute-charge response of scintillating screens was calibrated against the ELBE accelerator in a separate campaign~\cite{Kurz2018}. 

\subsection{Particle-in-cell simulations}

The three-dimensional start-to-end simulations shown in Fig.~\ref{fig:PIC-sim} were performed with the particle-in-cell code PIConGPU ~\cite{Bussmann2013, Burau2010}, version 0.4.2~\cite{Huebl2018}. 
The simulation closely approximates the experimental parameters of the pre-ionized LPWFA by modeling the measured transverse laser pulse profile by including higher Laguerre-Gauss laser modes as well as modeling the measured gas density and gas mixture used in both the LWFA and PWFA stages. 
As in the experiment, a foil is inserted after the PWFA up-ramp to reflect the laser. For this purpose, the simulated foild was implemented with 50 times the critical density, which is sufficiently dense to lead to the laser-plasma mirror effect. This approach neglects density perturbations around the foil and possibly underestimates the divergence increase due to fields within the foil, as observed in the experiments and also reported in~\cite{Raj2019}. 
Closely mimicking the experiment, the central wavelength of the laser is 800 nm, the total energy 1.4 J, the pulse duration 30 fs, and the spot size \SI{19}{\micro\meter} (both FWHM intensity). 
The moving-window frame of the simulation has a total size of $768\times\,4608\times\,768$ cells and propagates for 300,000 iterations. The spatial resolution is 177 $\times\,$44.3 $\times\,$177 nm with a temporal resolution of 72.8 as. 
The electromagnetic field evolution is simulated with the Lehe solver~\cite{Lehe2013} including a binomial filter~\cite{Birdsall2005}, while the particle motion is computed using the Boris pusher~\cite{Boris1970}. Particles influence the fields via the Esirkepov current deposition scheme~\cite{Esirkepov2001} with a TSC macro-particle shape~\cite{Hockney1988}. 
Ionization was treated via a combined BSI~\cite{Bauer1999} and ADK~\cite{Delone1998} model. The complete input data set is available online~\cite{Richard2019}.

Simulations used for the illustration of the concept in Fig.~\ref{fig:setup}(b) and (c) were performed using the code OSIRIS~\cite{Fonseca2002}. 

\subsection{Dual-bunch experiment}
The dual-bunch LPWFA experiment was carried out with the ATLAS Ti:Sa laser system. 
In this experiment, two supersonic de Laval nozzles and a shock-front injector~\cite{Buck2013} were used. 
The shock was created by the edge of a silicon wafer partially obstructing the gas flow in the first jet. 
The outlet diameters and Mach numbers of the first (LWFA) and second (PWFA) nozzle were \SI{5}{\milli\meter}, $M = 6.35$ and \SI{1}{\milli\meter}, $M = 5$, respectively, with a vacuum-gap of \SIrange{5.5}{6}{\milli\meter}. 
In the LWFA stage, the dual-energy bunch pair is generated via shock-front injection into the first and second period of the plasma wakefield. 
The double-bunch distribution, forming the driver and witness, are hence separated by approximately one plasma wavelength $\lambda_p\,\simeq \SI{17}{\micro\metre}$ at a density of $\SI{4.0e18}{\per\cubic\centi\metre}$. Supplementary Fig.~5 shows the interferometrically measured electron density profile along the interaction axis. The PWFA stage was operated at $\SI{1.4e18}{\per\cubic\centi\metre}$. The electron energy was characterized using a \SI{80}{\centi\meter}-long, \SI{0.8}{\tesla} permanent dipole magnet spectrometer, resolving energies from \SI{50}{\mega\electronvolt} onward.

\newpage
\begin{description}
 \item [Acknowledgments]
This project was fully supported by the Helmholtz association under program Matter and Technology, topic Accelerator Research and Development. The authors gratefully acknowledge the GWK support for funding this project by providing computing time through the Center for Information Services and HPC (ZIH) at TU Dresden on the HRSK-II. A.M.O. thanks the OSIRIS consortium (IST/UCLA) for access to the OSIRIS code, acknowledges the grant of computing time by the J{\"u}lich Supercomputing Center on JUWELS under Project No. HHH45 and the use of the High-Performance Cluster (Maxwell) at DESY.
 M.F.G, H.D., A.D{\"o}., J.G., S.Schi. and S.K. acknowledge support by the DFG through the Cluster of Excellence Munich--Centre for Advanced Photonics (MAP EXC 158), by Euratom research and training programme under Grant Agreement No. 633053 within the framework of the EUROfusion consortium, and the Max Planck Society. 
T.H., A.M.O and R.W.A. acknowledge support by the European Union's Horizon 2020 research and innovation programme under Grant Agreement No. 653782 (EuPRAXIA).
 S.C., O.K. and G.R. were supported by the European Research Council (ERC) under the European Union's Horizon 2020 research and innovation programme (Miniature beam- driven Plasma Accelerators project, ERC Grant agreement No. 715807).
 \item [Author' contributions]
T.K., T.H., J.P.C.C., O.K., S.S., R.W.A., S.C., B.H., U.S., A.M.O. and A.I. planned and designed the high-gradient experiment.
T.K., T.H., M.F.G., Y.Y.C., J.P.C.C., O.K., S.S., J.G., A.K., O.Z. and A.I. contributed to experimental work.
T.K., T.H., J.P.C.C., O.K., S.S., A.M.O. and A.I. analyzed the data.
R.P., A.D., K.S., M.B. and A.M.O. performed the numerical simulations;
M.F.G., H.D., J.G., S.Schi., A.D{\"o}. and S.K. designed and carried out the dual-bunch experiment.
M.F.G., A.D{\"o}. and S.K. analyzed the data.
J.G. and A.D{\"o}. performed numerical simulations;
T.K., T.H., M.F.G., O.K., Y.Y.C., J.P.C.C., A.D., O.K., R.P., S.S., R.W.A., G.R., S.C., A.Dö., B.H., S.K., U.S., A.M.O. and A.I. discussed all experimental and numerical results.
All authors contributed to writing the manuscript.
 
 \item [Additional information]
 \item [Correspondence] Correspondence and requests for materials should be addressed to T.K.~(email: t.kurz@hzdr.de), A.I.~(email: a.irman@hzdr.de) and S.K.~(email: stefan.karsch@mpq.mpg.de)
 \item [Data availability]
 The data that support the figures and further findings of this article are available from the corresponding authors upon reasonable request. 
 \item [Competing Interests] The authors declare no competing financial interests.
\end{description}

\section*{References}

%

\end{document}